\documentclass{elsart}

\usepackage{natbib}

 \usepackage{epsfig}

\usepackage{psfrag}
\usepackage{amsfonts}          
\usepackage{amssymb}           
\usepackage{amsmath}           
\usepackage{color}

\renewcommand{\vec}[1]{\mbox{\boldmath $ #1$}}

\newcommand{\rev}[1]{#1}

\begin{document}

\begin{frontmatter}



\title{Toroidal Flux Oscillations as Possible Causes of Geomagnetic Excursions and Reversals}


\author{F. H. Busse}

\address{Institute of Physics, University of Bayreuth, D-95440 Bayreuth,  Germany}
\ead{busse@uni-bayreuth.de}

\author{R. D. Simitev}

\address{Department of Mathematics, University of Glasgow, G12 8QW Glasgow, UK}
\ead{r.simitev@maths.gla.ac.uk}
\begin{abstract}
It is proposed that convection driven dynamos operating in planetary
cores could be oscillatory even when the oscillations are not directly
noticeable from the outside. Examples of dynamo simulations are
pointed out
that exhibit oscillations in the structure of the azimuthally averaged
toroidal magnetic flux while the mean poloidal field shows only
variations in its amplitude. In the case of the geomagnetic field,
global excursions may be associated with these oscillations.
Long period dynamo simulations indicate that the oscillations
may cause
reversals once in a while. No special attempt has been made to use
most realistic parameter values. Nevertheless some similarities
between the simulations and the paleomagnetic record can be pointed
out.
\end{abstract}


\end{frontmatter}

\section{Introduction}

The origin of geomagnetic reversals is a much debated subject among
scientists in the fields of paleomagnetism and dynamo theory. There is
general agreement that a detailed understanding of reversals is a key
issue of geodynamo theory. In this connection also the
problem of global excursions of the geomagnetic field in which the
dipole strength reaches temporarily unusually low values has been
discussed and it has been suggested (\cite{DC72}, \cite{H81}, see
also \cite{M96} and later papers by \cite{G99} and \cite{W}) that global
excursions are aborted reversals. Not all recorded excursions are
global ones, but global excursions still appear to occur more
frequently than reversals. \cite{L97} identified at least six global
excursions in the last about 800 ky since the Brunhes/Matuyama
reversal, but in the  last decade many more global excursions have
been found according to \cite{Lu2}. In this letter we wish  
to support the notion that global excursions and reversals originate
from the same mechanism. An oscillatory dynamo process that manifests
itself primarily in the toroidal component of the magnetic field will
be proposed as such a mechanism. 
\rev{Indeed, from the perspective of the oscillations, excursions must be considered as the normal behavior, while a reversal represents an exceptional excursion in which the  mean poloidal field is perturbed more strongly that it can recover from its low-amplitude state only with the opposite sign. }

Traditionally the geodynamo is regarded as a stationary dynamo in
contrast to the solar dynamo which exhibits a 22-year period. Dynamo
simulations have shown, however, that in rapidly rotating spherical fluid
shells with significant differential rotation often oscillatory
dynamos are found.  That
dynamo oscillations may not be visible from the exterior of the
conducting fluid sphere has been pointed out previously (\cite{BS06}). The present letter intends to
demonstrate how oscillations can lead to global excursions and more
rarely to reversals. While the simulations are based on the fundamental
equations governing the generation of magnetic fields by convection
flows in rotating spherical shells, only a minimum of physical
parameters is introduced and a faithful modeling
of the Earth's core has not been the primary goal.

\section{Mathematical formulation}

We consider a  spherical fluid shell of thickness $d$ rotating with a constant angular
velocity $\Omega$. It is assumed that a static state exists with the
temperature distribution $T_S = T_0 -\beta d^2 r^2 /2$. Here $rd$ is
the length of the position vector, \vec{r}, with respect to the center of the
sphere. The gravity field is $\vec g = - d \gamma \vec r$. In addition
to the length $d$, the time $d^2 / \nu$,  the temperature $\nu^2 /
\gamma \alpha d^4$ and the magnetic flux density $\nu ( \mu \varrho
)^{1/2} /d$ are used as scales for the dimensionless description of
the problem  where $\nu$ denotes the kinematic viscosity of the fluid,
$\kappa$ its thermal diffusivity, $\varrho$ its density, $\alpha$ 
its coefficient of thermal expansion and $\mu$ is its magnetic
permeability.  The Boussinesq approximation is assumed.
Accordingly, the velocity field $\vec u$ as well as the magnetic flux
density $\vec B$ are solenoidal vector fields for which the general
representation in terms of poloidal and toroidal components can be
used, 

\vspace{-1.5cm}        
\begin{subequations}
\begin{gather}
\vec u = \nabla \times ( \nabla v \times \vec r) + \nabla w \times
\vec r \enspace, \\
\vec B = \nabla \times  ( \nabla h \times \vec r) + \nabla g \times
\vec r \enspace.
\end{gather}
\end{subequations}

\vspace{-0.5cm}\noindent 
By multiplying the (curl)$^2$ and the curl of the
equation of motion and of the induction equation by  
$\vec r$, we obtain four equations for $v$ and $w$ and for $h$ and 
$g$. These four equations together with the heat equation for the
dimensionless deviation $\Theta$ from the static temperature
distribution and with the appropriate boundary conditions represent
the basis for the mathematical description of the evolution in time of
thermal convection in the rotating spherical shell and of the magnetic
field generated by it. Since these equations have been given in
previous papers (\cite{SB05}, \cite{BS06}), we list here only the
dimensionless parameters, the Rayleigh number $R$, the Coriolis number
$\tau$, the Prandtl number $P$ and the magnetic Prandtl number $P_m$, 

\vspace{-1cm}
\begin{equation}
R = \frac{\alpha \gamma \beta d^6}{\nu \kappa} ,
\enspace \tau = \frac{2
\Omega d^2}{\nu} , \enspace P = \frac{\nu}{\kappa} , \enspace P_m =
\frac{\nu}{\lambda},
\end{equation}

\vspace{-0.5cm}\noindent 
where $\lambda$ is the magnetic diffusivity. We assume stress-free
boundaries with fixed temperatures and use the radius ratio
$r_i/r_o=0.4$, 

\vspace{-1.5cm}
\begin{gather}
\hspace{-15mm}
 v = \partial^2_{rr}v = \partial_r (w/r) = \Theta = 0 \nonumber \\
\hspace{20mm} 
 \mbox{ at } r=r_i \equiv 2/3  \mbox{ and } r=r_o \equiv 5/3.
\label{vbc}
\end{gather}

\vspace{-0.5cm}\noindent 
For the magnetic field an electrically insulating outer
boundary is assumed such that the poloidal function $h$ must be
matched to the function $h^{(e)}$ which describes the
potential field outside the fluid shell

\vspace{-1.5cm}
\begin{gather}
\hspace{-15mm}
g = h-h^{(e)} = \partial_r ( h-h^{(e)})=0 \: \mbox{ at }  r=r_o \equiv 5/3.
\label{mbc1}
\end{gather}

\vspace{-0.5cm}\noindent
In order to avoid the computation of $h$ and $g$ in the inner core, $r
\le r_i$, we assume either an electrically insulating inner boundary, 

\vspace{-1.5cm}
\begin{gather}
\hspace{-15mm}
g = h-h^{(e)} = \partial_r ( h-h^{(e)})=0\:
\mbox{ at } r=r_i \equiv 2/3,
\label{mbc1}
\end{gather}

\vspace{-0.5cm}\noindent 
or a perfectly conducting inner core in which case the conditions 

\vspace{-1cm}
\begin{equation}
 h = \partial_r(r g) = 0 
\qquad \mbox{ at } r=r_i \equiv 2/3
\end{equation}

\vspace{-0.5cm}\noindent
must be applied. The numerical integration of the equations
  for $v, w, \Theta, h$ and $g$   together with
boundary conditions (3), (4)  and (5) or (6) 
proceeds with the pseudo-spectral method as described by \cite{T}
which is based on an expansion of all dependent variables 
in spherical harmonics for the $\theta , \varphi$-dependences, i.e. 

\vspace{-1.5cm}
\begin{equation}
h = \sum \limits_{l,m} H_l^m (r,t) P_l^m ( \cos \theta ) \exp \{ im
\varphi \}
\end{equation}

\vspace{-0.5cm}\noindent
and analogous expressions for the other variables, $v, w, \Theta$ and
$g$. $P_l^m$ denotes the associated Legendre functions. For the
$r$-dependence expansions in Chebychev polynomials are used. 
Azimuthally averaged components of the fields $v, w, \Theta, h$ and
$g$ will be indicated by an overbar. For most computations to be reported here
a minimum of 33 collocation points in the radial direction and
spherical harmonics up to the order 96 have been used. But this high
resolution was not needed in all cases. Instead of the time $t$
based on the viscous time scale we shall use in the following the
time $t^\ast = t/P_m$ based on the magnetic diffusion time,
$d^2/\lambda$. 

\vspace{-0.5cm}
\section{Oscillations of the Toroidal Magnetic Flux}
Even in their turbulent state of motion, convection flows outside the
tangent cylinder which touches the inner core boundary at its equator
remain essentially symmetric with respect to equatorial plane as is
evident from figure 1. For this reason dynamo solutions characterized
by an axial dipole correspond to a mean azimuthal magnetic flux that
is antisymmetric with respect to the equatorial plane. Oscillations of
these axisymmetric flux tubes originate from the creation of a pair of
new flux tubes with opposite signs at the equatorial plane which grow
and push the older flux towards higher latitudes as shown in figure
2. This process is strongly dependent on the differential rotation
which is prograde at larger distances from the axis and retrograde at
smaller ones. The oscillations can be described by Parker's dynamo wave model
(\cite{P}) as has been done by \cite{BS06}. 
\rev{In the present case of figure 2 the oscillation is modified in two respects. First, the mean toroidal field becomes nearly quadrupolar, i.e. symmetric about the equatorial plane, as the amplitudes of the mean poloidal field and of the differential rotation reach their minimum values. Secondly and more importantly, the mean poloidal field participates in the oscillation only as far as its amplitude varies. In the case of figure 2 its amplitude decays and
reaches a minimum around $t^\ast \approx 1.6$ at which} 
time a magnetic eddy emerges with the opposite sign of the given mean poloidal
field. Usually this eddy drifts outward and dissipates as it reaches
the surface of the conducting region such that the original poloidal
field prevails. Now a relatively long time passes before the process
repeats itself and new toroidal flux emerges at the equatorial
plane. In contrast to the thinner flux tubes of dynamos at higher
Prandtl numbers which exhibit a more sinusoidal oscillation as shown
in section (b) of figure 3, the oscillation in the
present case resembles more a relaxation oscillation as
shown in  section (a) of figure 3. The amplitudes $H_l^0$,
$G_l^0$ in this figure are assumed at the mid-radius of the fluid
shell, \rev{but $H_1^0$ usually does not differ much from the dipole component describing the magnetic field outside the fluid shell.}

While the process visualized in figure 2 shares several features
with global excursions, it may also give rise to reversals. These
happen in some cases when the emerging eddy with the opposite sign of
the poloidal field replaces the latter as shown in figure 4. 
This situation occurs most likely if the eddy with the opposite sign
emerges near the equatorial plane such that it splits the original
field into two parts. It is remarkable that the reversed poloidal
field appears first at low latitudes as has also been observed in the
case of geomagnetic reversals (\cite{C}). Note that the radius
$r=r_o + 1.3$ corresponds approximately to the Earth's surface. 
\rev{The occurrence of a reversal seems to be promoted by a
  particularly strong equatorially symmetric  toroidal flux as appears
  to be indicated by the correlation between reversals and relative
  high absolute values of the coefficient $G_1^0$ in sections (b) and
  (c) of figure 3.
  We note in passing that Li et al. (2002), propose a reversal
  mechanism in which the quadrupole mode grows, exceeding the dipole
  mode before the reversal in a manner similar to what happens near
  the minimum of the oscillations shown in figure 2. In
  contrast, however, our dynamo solutions do not alternate between high- and
  low-energy states, nor do they exhibit a broken columnar vortex
  structure of the velocity field.}

The examples discussed so far all correspond to a single set of
parameter values. In particular condition (6) for a highly electrically
conducting core has been used. In order to demonstrate the robust
nature of the mechanism of global excursions and reversals, we show in
figure 5 a sequence of plots exhibiting a reversal from a dynamo
simulation with a quite different set of parameters for which 
condition (5) instead of (6) has been applied. The oscillations occur
somewhat less regularly in this case as is evident from the time series
of the amplitude of the axial dipole component shown in section (c) of
figure 3, \rev{but the average period is again close to half a magnetic diffusion time.}
A common property of the oscillations is that the quadrupolar
components of the axisymmetric magnetic field play a significant
role. In this respect some similarity may be noted with the
oscillations displayed in figures 12 and 13 of \cite{BS06}. 

Although the inner core does not participate in the oscillations
in either of the boundary conditions (5) and (6), we expect that the
use of a vanishing jump of the electrical conductivity at $r = r_i$
will not affect the results significantly. As has been observed in
the dynamo simulations of \cite{W2} and of \cite{SB05}, because of
its small volume the inner core does not appear to have a
significant effect on the dynamo process.

\vspace{-0.5cm}
\section{Discussion}

In selecting the dynamo cases displayed in figures 2, 4 and 5 we
have emphasized a high value of $\tau$ 
and a reasonably high value of $R$ for which the available computer
capacity allows to obtain time records extending over many
magnetic diffusion times. The critical values of the Rayleigh numbers
for $\tau = 3\times10^4$ and $\tau = 10^5$ are $R_c = 2.35\times10^4$
with $m_c=10$ and $R_c = 1.05\times10^6$ with $m_c=11$,
respectively. Hence the Rayleigh numbers used for the cases of figures
2, 4 and 5 exceed their critical values by nearly a factor of
four. The corresponding average Nusselt numbers at the inner
boundary are $Nu_i = 1.58$ and $Nu_i = 1.73$ 
\rev{ and the corresponding magnetic Reynolds numbers, defined by $R_m \equiv
  P_m\sqrt{2E}$, are $R_m=210$ and $R_m=156$, respectively}. 
The Prandtl number
$P = 0.1$ was used in both cases since it appears to be a reasonable
compromise between the molecular value $P = 0.05$ estimated for the
outer core (\cite{P88}) and a value of the order unity usually
assumed for a highly turbulent fluid. Moreover, the choice of a low
value of $P$ has allowed us to choose a desirable relatively low value of
$P_m$.  

\rev{The successful application of Parker's kinematic model for dynamo waves employed by \cite{BS06} suggests that the oscillations depend primarily on the differential rotation and the mean helicity of convection which are assumed as given. The modified oscillation considered in the present paper is characterized by an extended phase of a dominant equatorially symmetric (quadrupolar) mean toroidal field which is responsible for the property that the period becomes comparable to the magnetic diffusion time. The variations of the amplitude of convection and of the differential rotation seem to be of lesser importance.}
  
Using the depth $d\approx 2200$ km of the liquid outer core and a
typical and often quoted value $\lambda\approx 2$ m$^2$/s we find
$0.8\times10^5$ years as the magnetic diffusion time of the Earth's
core which corresponds to $t^\ast = 1$ in the figures of this
paper. The oscillation period $T^\ast \approx 0.5$ obtained 
in the time series of figure 3(a) thus roughly equals 
about 40 ky in the Earth's core. This period is quite
comparable to the broad maximum in the region of 30-50 ky that
seems to characterize the spectrum of the amplitude variations of the
geomagnetic field (\cite{TS}, \cite{TH}, \cite{GV}) throughout the
last million years. A more recent analysis   (\cite{CJ}) has shed some
doubts on the existence of such a spectral peak, but still confirms a
sharp  decrease of the spectral power for periods shorter than about
30 ky. We also like to draw attention to the  property that the
typical separation between global excursions in table 1 of \cite{Lu2}
varies between 30  and 50 ky.

From the reversals exhibited in figures 3, 4 and 5 it appears  
that the amplitude increases more sharply after the reversal than it
decays towards the reversal.
\rev{To demonstrate this effect more clearly we have plotted in figure 6
the coefficient $H_1^0$ in proximity of the reversal as a function of
time for each of the last 4 reversals that have been obtained in the
cases a) and c) of figure 3. Although the asymmetry between the
dipole strengths before and after the reversal is not as strong as
has been found in the case of  paleomagnetic reversals (\cite{V},
\cite{GV2}), a similar effect seems to exist. Since the time records of figure 3 do not exhibit this effect very well we have plotted in figure 6 values of $H_1^0$ at $r=r_i+0.5$ for shorter time periods. In the
case $R=850000$ $H_1^0$ at $r=r_o$ is also shown (by solid lines) since it represents the axial dipole strength of the potential field
outside the fluid shell. Apart from a small shift in time the value
of $H_1^0$ does not vary much as function of the radius.}
In the continuing investigation of the dynamo oscillations it will
be attempted to find even closer correspondences with paleomagnetic
observations.  

\rev{The possibility of toroidal flux oscillations as origin of global excursions and reversals proposed in this paper differs from all other mechanisms proposed in the literature for reversals and excursions and resembles more the mechanisms considered for the solar cycle. In the latter the mean poloidal field fully participates, of course, similarly as in the dipole oscillation of figure 10 of \cite{BS06} except for the property that the solar dynamo wave propagates towards lower instead of higher latitudes. A comparison of different mechanisms for geomagnetic reversals would go beyond the scope of present paper and should be postponed until more detailed  computational results for a wider range of parameters become available.}

\clearpage

\begin{figure}[t]
\begin{center}
\noindent\includegraphics[width=20pc]{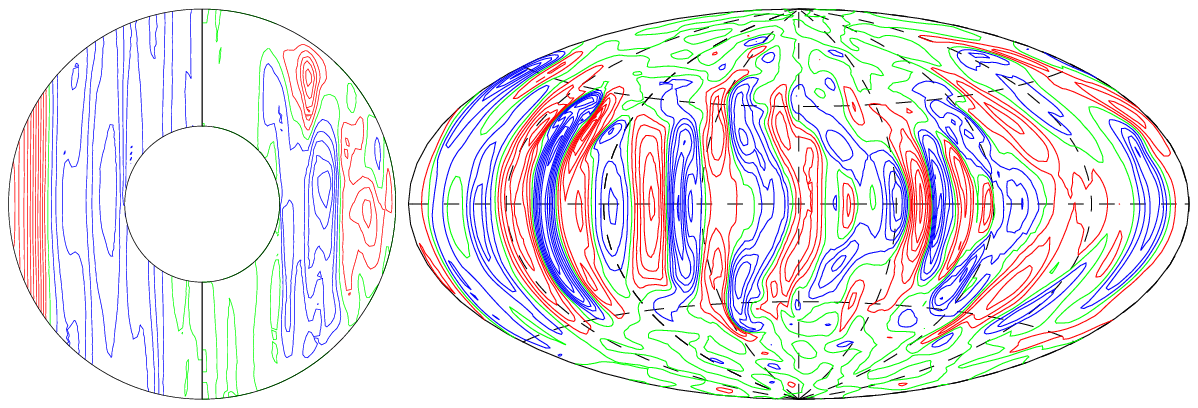}
\end{center}
\caption{(color online). Typical structures of the velocity field in the case $P=0.1$,
  $\tau=10^5$, $R=4\times10^6$, $P_m=0.5$ with a perfectly electrically conducting
inner core. The left plot shows 
lines of constant $\overline{u}_\varphi$ in the left half and
streamlines $r \sin\theta \partial_\theta \overline{v}=$const.~in the right
half, all in the meridional plane. The right plot shows lines of
constant $u_r$ at $r=r_i+0.5$ at the time $t^\ast=1.486$. Positive
and negative values are indicated by solid (red online) and dashed
(blue online) lines.} 
\end{figure}

\clearpage

\begin{figure}[t]
\begin{center}
\noindent\includegraphics[width=26pc]{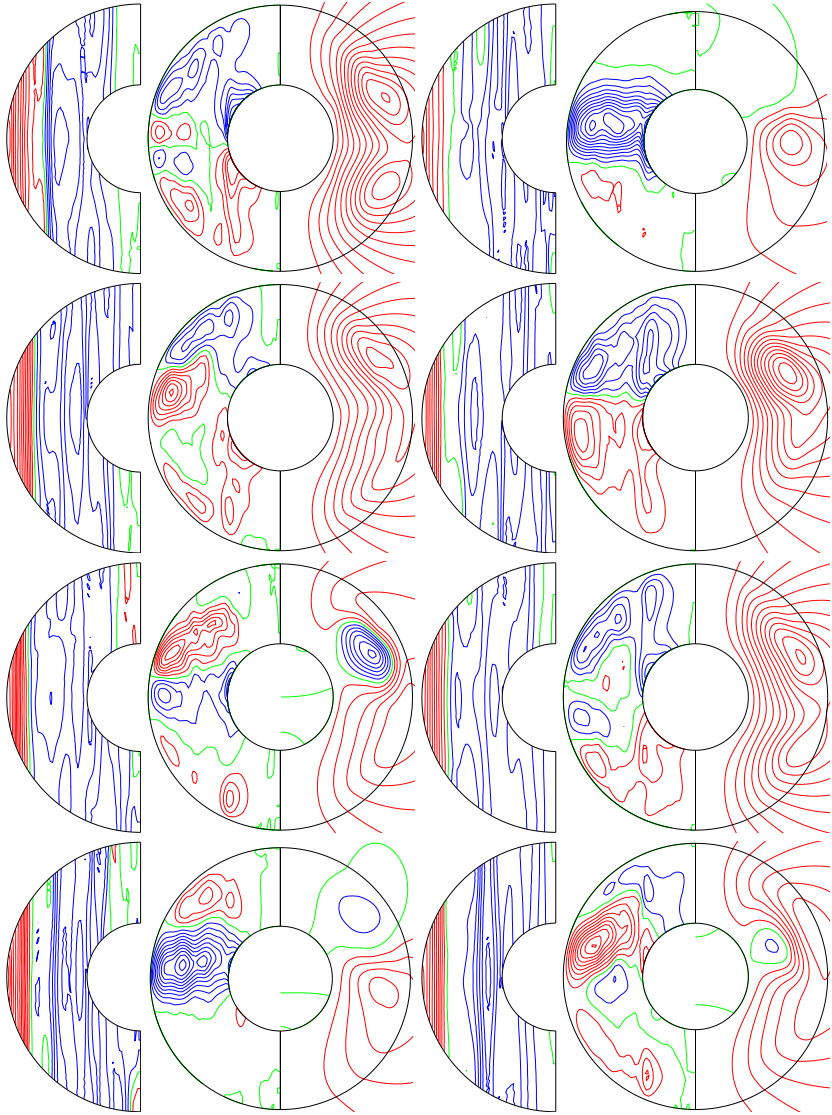}
\end{center}
\caption{(color online). Dynamo oscillation in the case  $P=0.1$,
  $\tau=10^5$, $R=4\times10^6$, $P_m=0.5$ with perfectly conducting
  inner core. \rev{The half circles show lines of constant
  $\overline{u}_\varphi$.} The full circles show meridional
  isolines of  $\overline{B}_\varphi$ (left half) and of $r\sin\theta
  \partial_\theta \overline{h}$ (right half) at times
$t^\ast=1.490$, $1.538$, $1.586$, $1.634$, (first column, from top to
  bottom) and  
$t^\ast=1.682$, $1.810$, $1.954$, $2.034$ (second column).
The times $t^\ast$ refer to figure 3(a).}
\end{figure}

\clearpage

\begin{figure}[t]
\begin{center}
\noindent\includegraphics[width=19pc,angle=0,clip]{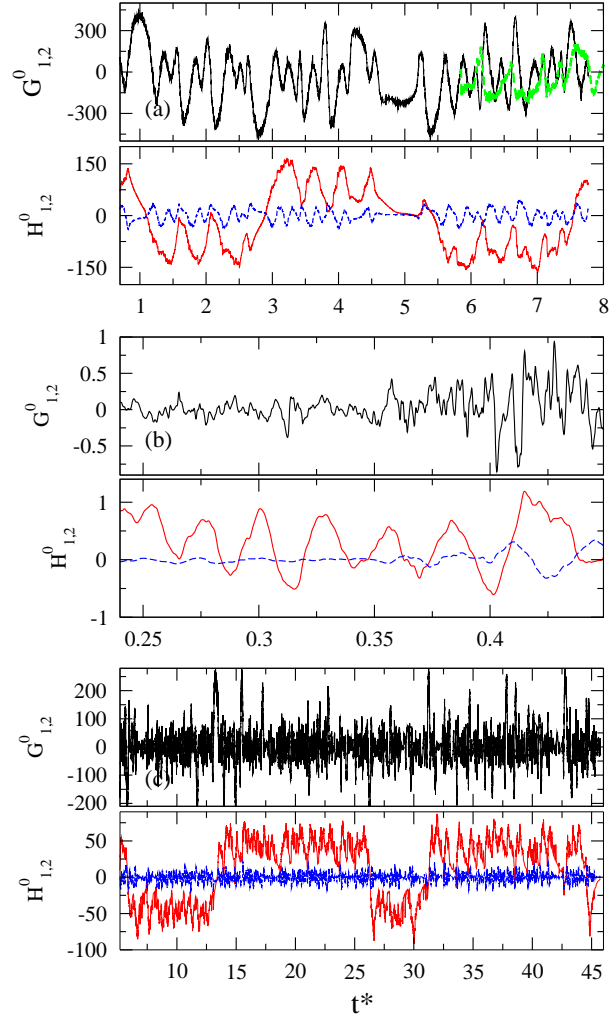}
\end{center}
\caption{(color online). Selected coefficients at $r=r_i+0.5$ in the cases 
 (a) $P=0.1$, $\tau=10^5$, $R=4\times10^6$, $P_m=0.5$ with perfectly
 conducting inner core;
 (b)  $P=5$, $\tau=5000$, $R=600000$, $P_m=10$ with electrically insulating inner core;
 (c)  $P=0.1$, $\tau=3\times10^4$, $R=850000$, $P_m=1$ with insulating inner core.
 The coefficient of the axial dipole component $H^0_1$ (axial
 quadrupole component $H^0_2$) is indicated by a solid/red online
 (dashed/blue online) line. \rev{The coefficient $G^0_2$ in (a) is
 indicated by a dashed/green online line.}}  
\end{figure}

\clearpage

\begin{figure}[t]
\begin{center}
\noindent\includegraphics[width=20pc]{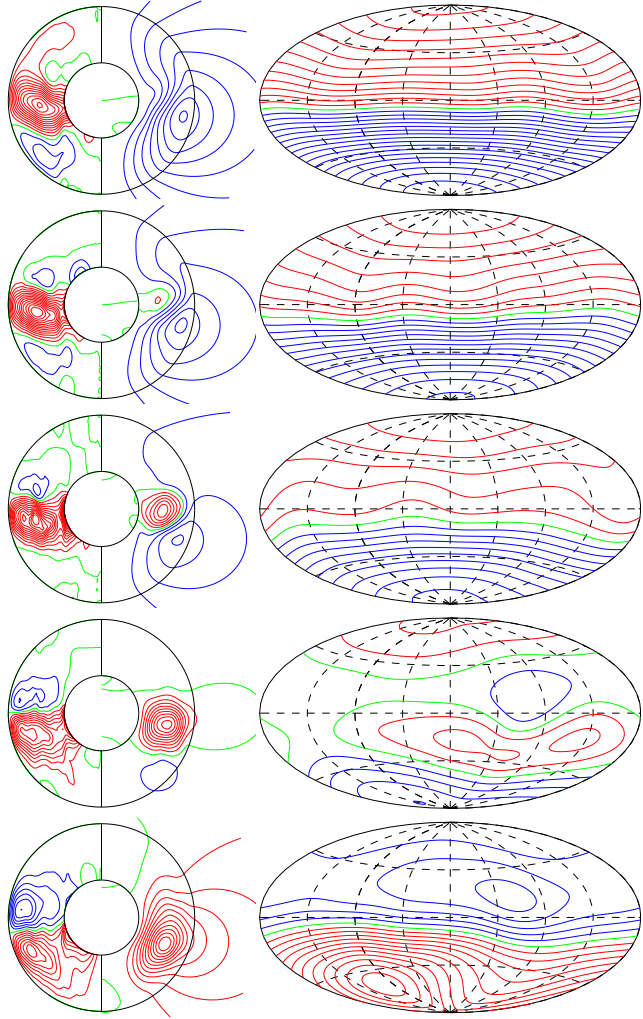}
\end{center}
\caption{(color online). Magnetic field polarity reversal in the case  $P=0.1$,
  $\tau=10^5$, $R=4\times10^6$, $P_m=0.5$ with perfectly conducting
  inner core. The left column shows meridional
  isolines of  $\overline{B}_\varphi$ (left half) and of $r\sin\theta
  \partial_\theta \overline{h}$ (right half). The right column shows
  lines $B_r=$ const.\  at $r=r_o+1.3$. The interval
  between the plots is $\Delta t^\ast=0.048$ with the first plot at
  $t^\ast=0.994$ (see figure 3(a)).}
\end{figure}

\clearpage

\begin{figure}[t]
\begin{center}
\noindent\includegraphics[width=20pc]{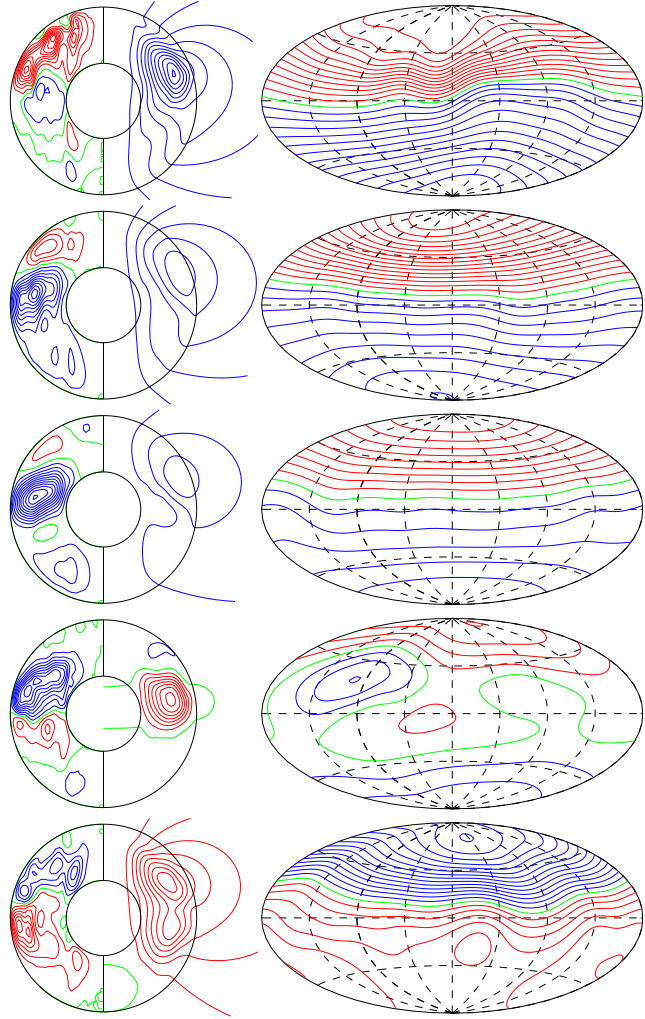}
\end{center}
\caption{(color online). Same as figure 4, but for $P=0.1$,
  $\tau=3\times10^4$, $R=850000$, $P_m=1$ with insulating inner core.
   The interval between the plots is $\Delta t^\ast=0.07$ with
  the first plot at $t^\ast=26.155$ (see figure 3(c)).}
\end{figure}

\clearpage

\begin{figure}[t]
\begin{center}
\noindent\includegraphics[width=19pc,angle=0]{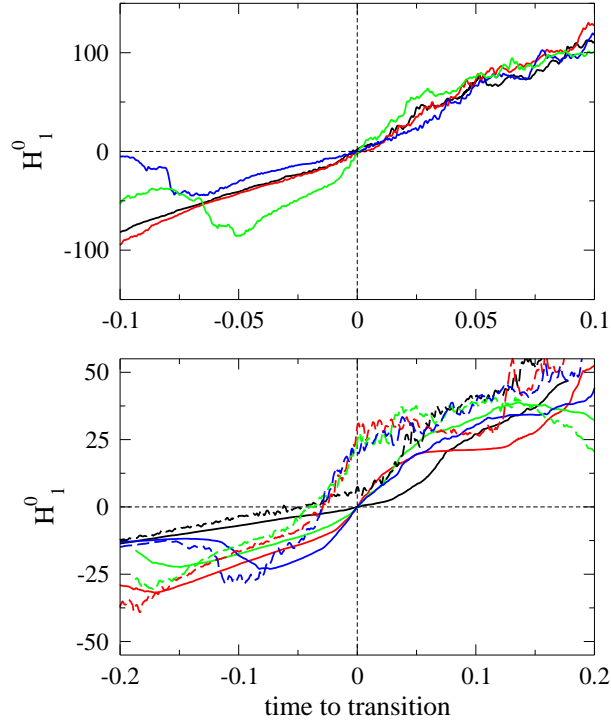}
\end{center}
\caption{(color online). Time-series of the coefficients of the axial dipole component
  $H^0_1$ at $r=r_i+0.5$  across the last 4 reversals in the cases
  $P=0.1$, $\tau=10^5$, $R=4\times10^6$,
  $P_m=0.5$ with perfectly  conducting inner core (top) and  $P=0.1$,
  $\tau=3\times10^4$, $R=850000$, $P_m=1$  with insulating inner core 
  (bottom). For the sake of comparison, the time series have been
  translated along the time axis so that the polarity transitions
  occur at $t=0$ and $-H^0_1$ is plotted for every second reversal. In
  both panels, black, red, blue and green color correspond to
  reversals 1(2) to 4(5) of the respective cases in figure 3. \rev{
  In the bottom panel, $H^0_1$ at $r=r_o$ has been included  in order to represent the axial dipole strength of the potential field
outside the fluid shell.  $H^0_1$ at $r=r_i+0.5$ (given by dashed lines) precedes it by about $\Delta t^\ast\approx0.04$}}
\end{figure}

\end{document}